# GHZ-Equivalent State Distribution in Quantum Networks: Reducing Decoherence and Quantum Resource Consumption

Chun-Hsiang Wang[1, a] and Chia-Wei Tsai[1, *]

[1]Department of Computer Science and Information Engineering, National Taichung University of Science and Technology, No.129, Sec. 3, Sanmin Rd., North Dist., Taichung 40401, Taiwan.

[a]jam930725@gmail.com
[*]cwtsai@nutc.edu.tw

## Abstract

This study proposes a novel scheme for distributing GHZ-equivalent states across repeater-based quantum networks, with particular focus on the analysis and mitigation of decoherence effects during transmission. The proposed scheme enables remote users to share graph states, which can be leveraged to implement various quantum communication protocols, such as quantum key distribution and quantum secret sharing. Compared with existing approaches, the proposed distributed scheme requires only $O(N)$ qubits without introducing redundant entanglement structures. Together with the linear-scaling merging procedure in both controlled gate count and qubit usage, the proposed framework supports more efficient large-scale graph state distribution. To evaluate its feasibility and correctness, this study utilizes the quantum network simulation tool, NetSquid, to implement the proposed scheme. Simulation results demonstrate that the proposed approach is both effective and practical for executing quantum communication protocols within quantum networks.

## 1. Introduction

As quantum computing has advanced in recent years, studies have revealed the potential threats it poses to classical cryptographic systems. For instance, Shor's algorithm [1] can factor large integers in polynomial time, which will break the RSA scheme [2] and compromise the security of daily digital communications. In response to these threats, one promising solution is to develop quantum communication protocols (QCPs) based on the principles of quantum mechanics. Over the past decades, various types of QCPs have been proposed, including quantum key distribution (QKD) protocols [3-5], which enable two participants to share a secret key; quantum secret sharing (QSS) protocols [6-8], which allow a dealer to distribute a secret among multiple agents such that only specific subsets can reconstruct it; and quantum secure direct communication (QSDC) protocols [9,10], which enable direct information transmission over quantum channels without the need for pre-shared keys. Many of these protocols have been further studied and extended to accommodate different application scenarios, such as multiparty, lightweight, and semi-quantum environments [4,5,7,8].

To enable remote parties to perform quantum communication, many QCPs rely on the use of entangled states (e.g., GHZ states [11], W states [12], and graph states [13]) and often assume the existence of direct quantum channels between participants. However, in practical implementations, the assumption is difficult to satisfy because entangled qubits cannot be transmitted over long distances with high fidelity. To address this issue, quantum networks (QNs) [14] based on quantum repeaters can be employed to facilitate reliable transmission between two remote parties or the distribution of entanglement states among multiple parties.

Despite the difficulty of storing and transmitting qubits while maintaining high fidelity, additional challenges arise when distributing entangled states over quantum networks. Most quantum networks exhibit topological constraints, dynamic link conditions, and limited quantum memory resources [15], making the aforementioned assumptions hard to achieve in practice. Therefore, how to efficiently generate and distribute entangled states over quantum networks has become a critical research issue. Many studies [16-23] with different objectives have been conducted on generating and distributing entangled states over quantum networks. However, although these works pursue different objectives, none explicitly focuses on analyzing and mitigating decoherence during transmission. To address this issue, this study adopts the principle of minimizing quantum information exposure to noise, that is, mitigating the impact of coherence, by reducing the number of qubits transmitted, the transmission distances, and the duration qubits must be preserved. Given the significant impact of decoherence on the correctness and performance of QCPs, this study leverages the properties of graph states to propose a scheme for generating and distributing star graph states over quantum repeater-based quantum networks. Star states, which are LOCC-equivalent to GHZ states and can also be transformed into complete graph states, receive primary focus in this study because they provide strong multipartite entanglement that is particularly suitable for QCPs in quantum network. These states enable quantum key distribution [5], quantum secret sharing [8], quantum summation [24], and other communication protocols [25-27], where a single central qubit can efficiently coordinate the entanglement shared among multiple participants. The proposed method is specifically designed to minimize the impact of decoherence during transmission while also reducing the number of qubits and quantum operations required. Specifically, the contributions of this study include:

1. A distribution scheme for GHZ-equivalent states is proposed, which can serve as a building block for quantum network services. The proposed scheme leverages properties of graph states and aims to mitigate decoherence during transmission.
2. Compared with previous methods, the proposed scheme achieves linear scaling in both qubit usage and controlled gate count, supporting more efficient and reliable graph state distribution.
3. This study implements the proposed scheme using the quantum network simulation tool NetSquid [28] to verify its correctness and to evaluate the impact of quantum channel noise on the success rate.

The remainder of this paper is organized as follows. **Section 2** introduces the properties and operations of graph states, and reviews related works. **Section 3** details the proposed scheme for distributing target graph states. **Section 4** presents simulation results, comparisons and evaluations. Finally, **Section 5** concludes the paper and discusses potential directions for future research.

## 2. Preliminaries

In this section, we first introduce the definition of graph states and the associated operations derived from their properties, which are utilized in the proposed scheme. Next, we describe the characteristics of star and complete graphs. Then, we discuss the reduction in required operations and the corresponding corrections that must be applied following certain procedures. Finally, we review the related works of entanglement distribution methods.

### 2.1 Graph state

A quantum graph state corresponds to a graph $G = (V, E)$, where $V$ and $E$ denote the sets of vertices and edges, respectively. The corresponding quantum state can be formalized as:

$$|G\rangle = \prod_{(a,b)\in E} CZ^{\{a,b\}} \left(|+\rangle^{\otimes n}\right), \tag{1}$$

where $n = |V|$ is the number of vertices. Each vertex is initialized in the state $|+\rangle = \frac{1}{\sqrt{2}}(|0\rangle + |1\rangle)$. For each edge connecting two vertices $a$ and $b$, a controlled-Z (CZ) operation (shown in **Eq. (2)**) denoted as $CZ^{\{a,b\}}$ , is applied to the corresponding qubits.

$$CZ = \begin{bmatrix} 1 & 0 & 0 & 0 \\ 0 & 1 & 0 & 0 \\ 0 & 0 & 1 & 0 \\ 0 & 0 & 0 & -1 \end{bmatrix} \tag{2}$$

In this work, we use the local complementation (LC) operation and the Y-basis measurement introduced in [13] to propose a novel distribution method for GHZ-like states.

### 2.2 Star and Complete Graphs

A star graph is defined as a central node connected to all other leaf nodes, with no entanglement existing between any pair of leaf nodes. A star graph state $|G\rangle$ consisting of $n$ qubits, with $q_1$ as the central node, can be expressed as **Eq. (3)**, where $j_{(2)}$ denotes the binary representation of $j$. A star topology is a common communication network structure [29], and its corresponding entangled state can be leveraged for designing QCPs [5]. It is worth noting that a star graph with three or more nodes is equivalent to a GHZ state, as a star graph state can be transformed into a GHZ state using local operations and classical communication (LOCC).

$$|G\rangle = \prod_{2\le i\le n} CZ^{\{1,i\}} \sum_{j=0}^{2^n-1} \frac{1}{\sqrt{2}^n} \left|j_{(2)}\right\rangle_{12\ldots n} = \frac{1}{\sqrt{2}} (|0+\cdots+\rangle + |1-\cdots-\rangle)_{12\ldots n} \tag{3}$$

**Fig. 1**. An example of a 4-qubit star graph state.

A complete graph state $|G\rangle$ is a special type of graph state in which all nodes are pairwise connected, as expressed in **Eq. (4)**. Complete graph states find applications in various areas, such as QCP [8] and quantum repeaters [30].

$$|G\rangle = \prod_{1\le a<b\le n} CZ^{\{a,b\}}|+\rangle^{\otimes n} = \left(\frac{1}{\sqrt{2}}\right)^n \sum_{j=0}^{2^n-1} (-1)^{\Delta}|j\rangle, \tag{4}$$

where $\Delta = \sum_{0\le x<y\le n-1}\left(j_x \times j_y\right)$, and $j_x$ ($j_y$) denotes the $x$- (y-)th bit of $j$ in binary representation.

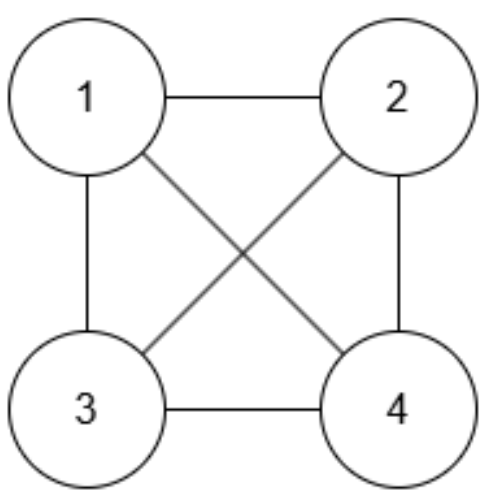

**Fig. 2**. An example of a 4-qubit complete graph state.

Star graph states and complete graph states can be transformed into one another through LC operations and single-qubit measurements [8,31]. Therefore, for simplicity, we primarily refer to the star graph state in the following discussion, with the understanding that complete graph states and GHZ states are implicitly included through such transformations.

### 2.3 Reductions on Required Operations and Corrections

The Clifford operators often take the form of $Z = \begin{bmatrix} 1 & 0 \\ 0 & -1 \end{bmatrix}, \sqrt{Z} = \begin{bmatrix} 1 & 0 \\ 0 & i \end{bmatrix}, Y = \begin{bmatrix} 0 & -i \\ i & 0 \end{bmatrix}, \sqrt{Y} = \frac{1}{2}\begin{bmatrix} 1+i & -1-i \\ 1+i & 1+i \end{bmatrix}$, and similar terms, which naturally arise after performing LC operations or Pauli measurements on graph states. According to Hein et al. [13], the relevant commutation relations between Pauli projection measurements and Clifford operations determine the operations and corrections applied to graph states. Moreover, certain corrections can be further simplified or omitted during these processes based on these commutation relations. **Table 1** summarizes the relationships that will be used in the following discussion.

**Table 1**. Commutation Relations between Pauli projection measurement and Clifford operators

| Pauli projection measurement | Clifford operator | Equivalent Result |
|---|---|---|
| $P_{X,\pm}$ | | $ZP_{X,\mp}$ |
| $P_{Y,\pm}$ | $Z$ | $ZP_{Y,\mp}$ |
| $P_{Z,\pm}$ | | $ZP_{Z,\pm}$ |
| $P_{Y,\pm}$ | $\sqrt{Z}$ | $\sqrt{Z}P_{X,\pm}$ |
| $P_{Z,\pm}$ | | $\sqrt{Z}P_{Z,\pm}$ |
| $P_{Y,\pm}$ | $\sqrt{Z}^{\dagger}$ | $\sqrt{Z}^{\dagger}P_{X,\mp}$ |
| $P_{Z,\pm}$ | | $\sqrt{Z}^{\dagger}P_{Z,\pm}$ |

Note that $\sqrt{Z}$ and $\sqrt{Z}^{\dagger}$ are equivalent to rotations $R_z\left(\frac{\pi}{2}\right)$ and $R_z\left(\frac{-\pi}{2}\right)$, respectively. The equivalent proofs are provided in **Appendix A**. For example, consider the operation $P_{Y,\pm}\sqrt{Z}^{\dagger}$**.** This corresponds to first applying $\sqrt{Z}^{\dagger}$ (i.e., $R_z\left(\frac{-\pi}{2}\right)$) to the qubit, followed by a measurement in the Y-basis, as shown in **Eq. (5)**, the overall effect is equivalent to measuring the qubit in the X-basis, with the measurement result inverted. This demonstrates that the original sequence of a rotation followed by a measurement can be replaced by a single measurement combined with classical post-processing, thereby reducing the required quantum operations, as shown in **Eq. (6)**. Detailed calculation procedures are given in **Appendix B**.

$$P_{Y,\pm}\sqrt{Z}^{\dagger} = \sqrt{Z}^{\dagger}P \tag{5}$$

$$P = \sqrt{Z}P_{Y,\pm}\sqrt{Z}^{\dagger} = P_{X,\mp} \tag{6}$$

### 2.4 Related Work

Here, we give a review on existing studies on entangled state distribution methods [16-23]. Ghaderibaneh et al. [16] proposed optimized entanglement swapping trees to improve long-distance entanglement generation in quantum networks. Chakraborty et al. [17] presented a linear programming framework based on multi-commodity flow to efficiently optimize entanglement distribution across quantum networks while meeting fidelity constraints. Ghaderibaneh et al. [18] conducted two efficient algorithms for generating and distributing high-fidelity GHZ states in quantum networks, showing superior performance via simulation. Fan et al. [19] presented a hypergraph-based linear programming framework to generate various types of graph states, achieving a high generation rate while considering the stochastic processes. Meignant et al. [20] used combinations of Bell pairs, local operations, and classical communication to propose protocols for distributing GHZ and arbitrary graph states across quantum networks. Koudia [21] proposed a Local Quantum Coding (LQC) strategy for distributing stabilizer states in quantum networks, enabling reductions in latency and memory overhead compared to conventional Bell-pair-based approaches. Both Fischer and Towsley [22] and Lin et al. [23] presented centralized graph state distribution methods for quantum networks, in which a central node locally constructs the desired graph state and then transfers the entanglement to the target nodes. However, although these works pursue different objectives, none explicitly focuses on analyzing and mitigating decoherence during transmission. Therefore, this study aims to propose a novel distributing scheme with a particular focus on analyzing and mitigating decoherence effects during transmission.

## 3. Proposed Scheme

In this section, we first introduce the distribution method along a path topology, followed by a discussion of the routing and root selection strategy employed in the scheme. We then describe the prototype process, which outlines the original step-by-step procedure underlying the proposed scheme. Finally, we present the optimized version of the scheme, derived from the prototype, and demonstrate its improved performance.

### 3.1 Distribution Along a Path

Based on the properties of graph states, the merging of two star graph states can be achieved by inserting an auxiliary qubit between the two central qubits and entangling it with each of them. Subsequently, sequentially Y-basis measurements on one central qubit and the auxiliary qubit, together with the corresponding corrections, transform the two into a single larger star graph state.

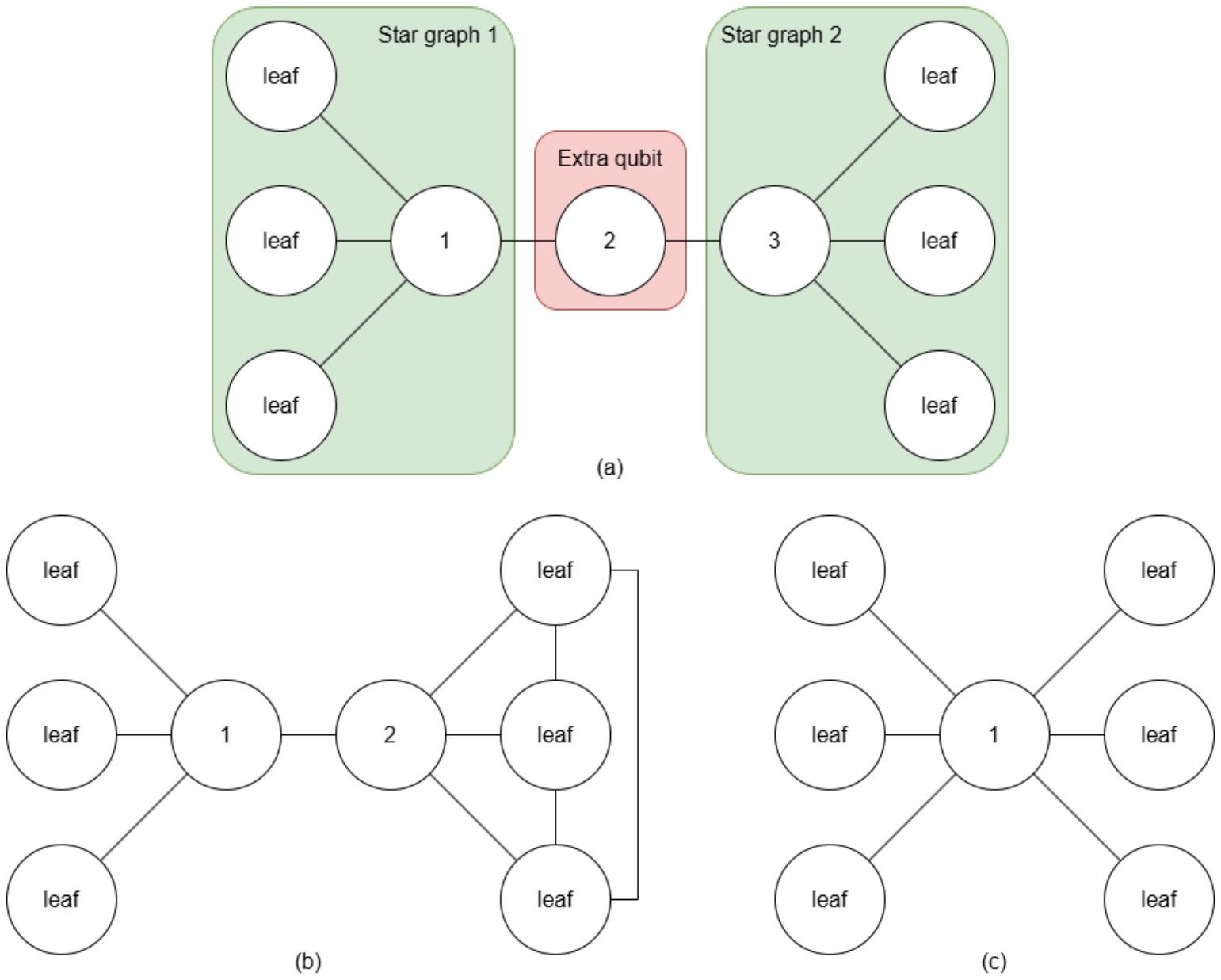

**Fig. 3**. The process of merging two star graph states. (a) shows the entangled state after inserting $q_2$. (b) illustrates the state after measuring $q_3$ in the Y-basis, (c) shows the final state after $q_2$ is measured in the Y-basis.

For two star graph states whose central qubits are connected by an intermediate chain of qubits, we can sequentially measure the qubits along the chain in the Y-basis until two central qubits are connected by a single qubit. Specifically, given a series of qubits, $\{q_1, q_2, \dots, q_n\}$, where $q_1$ and $q_n$ are the central qubits of different star graph states, we iteratively measure the qubit adjacent to $q_1$ in the Y-basis and apply the corresponding corrections until all qubits in $\{q_2, \dots, q_{\max(n-2,2)}\}$ have been measured, and the resulting entangled state can be reduced using the abovementioned method.

In this study, we employ a modified qubit distribution method based on [23] to merge two remote star graph states on the repeater-based quantum networks. Specifically, given a repeater chain $\{r_1, r_2, \dots, r_n\}$ where $r_n$ may be connected to end users, the process is carried out as follows:

**Step 1**: If $r_1$ or $r_n$ do not host a central qubit, then generate one qubit in the state $|+\rangle$ at that repeater.

**Step 2**: If $r_n$ is connected to end users, it generates one qubit for each end user and performs a CZ operation on the central qubit and each new qubit.

**Step 3**: If $n > 1$, perform the following steps:

(a) $r_1$ generates a qubit in the state $|+\rangle$, performs a CZ operation on its central qubit and the new qubit, then sends the new qubit to $r_2$.
(b) For each $r_i$ where $i = 2j + 1, j \in \{k | 1 \leq k < \lfloor n/2 \rfloor\}$, $r_i$ generates two qubits in the state $|+\rangle$, performs a CZ operation on them, then sends one to $r_{i-1}$ and the other to $r_{i+1}$.
(c) If $n \equiv 1 \pmod 2$ and $n > 2$, $r_n$ generates one qubit and performs a CZ operation on its central qubit and the new qubit, then sends the new qubit to $r_{n-1}$.

**Step 4**: For each $r_i$ where $i = 2j + 2, j \in \{k | 0 \leq k \leq \lfloor n/2 \rfloor - 1\}$, $r_i$ performs a CZ operation on the two qubits it holds, and measures the qubit received from $r_{i-1}$ in the Y-basis and applies corrections to adjacent qubits. The same process is then applied to the other qubit. This step is performed iteratively by each $r_i$.

**Step 5**: If $n \equiv 0\ (mod\ 2)$, perform the following steps:

(a) $r_n$ performs a CZ operation on its central qubit and the received qubit, then sequentially measures both the central qubit and the received qubit in the Y-basis, applying corrections to adjacent qubits.

Otherwise, proceed as follows:

(a) $r_{n-1}$ performs a CZ operation on the received qubits and measures the qubit from $r_{n-2}$ and applies corrections to adjacent qubits.
(b) $r_n$ measures its central qubit and applies corrections to adjacent qubits.
(c) $r_{n-1}$ then measures the qubit received from $r_n$ and applies corrections to adjacent qubits.

**Step 6**: If the whole process is finished, send qubits to the corresponding end users.

Note that even if the measured qubit and its adjacent qubits are held by different repeaters, the required corrections must still be applied across the relevant nodes. After this process, the two remote star graph states, originally centered at $r_1$ and $r_n$, are successfully merged into a larger star graph state centered at $r_1$. **Fig. 4** provides examples with four and five nodes in the repeater chain, respectively, illustrating the transformation achieved by the aforementioned method, and **Fig. 5** provides a brief flow chart for the process.

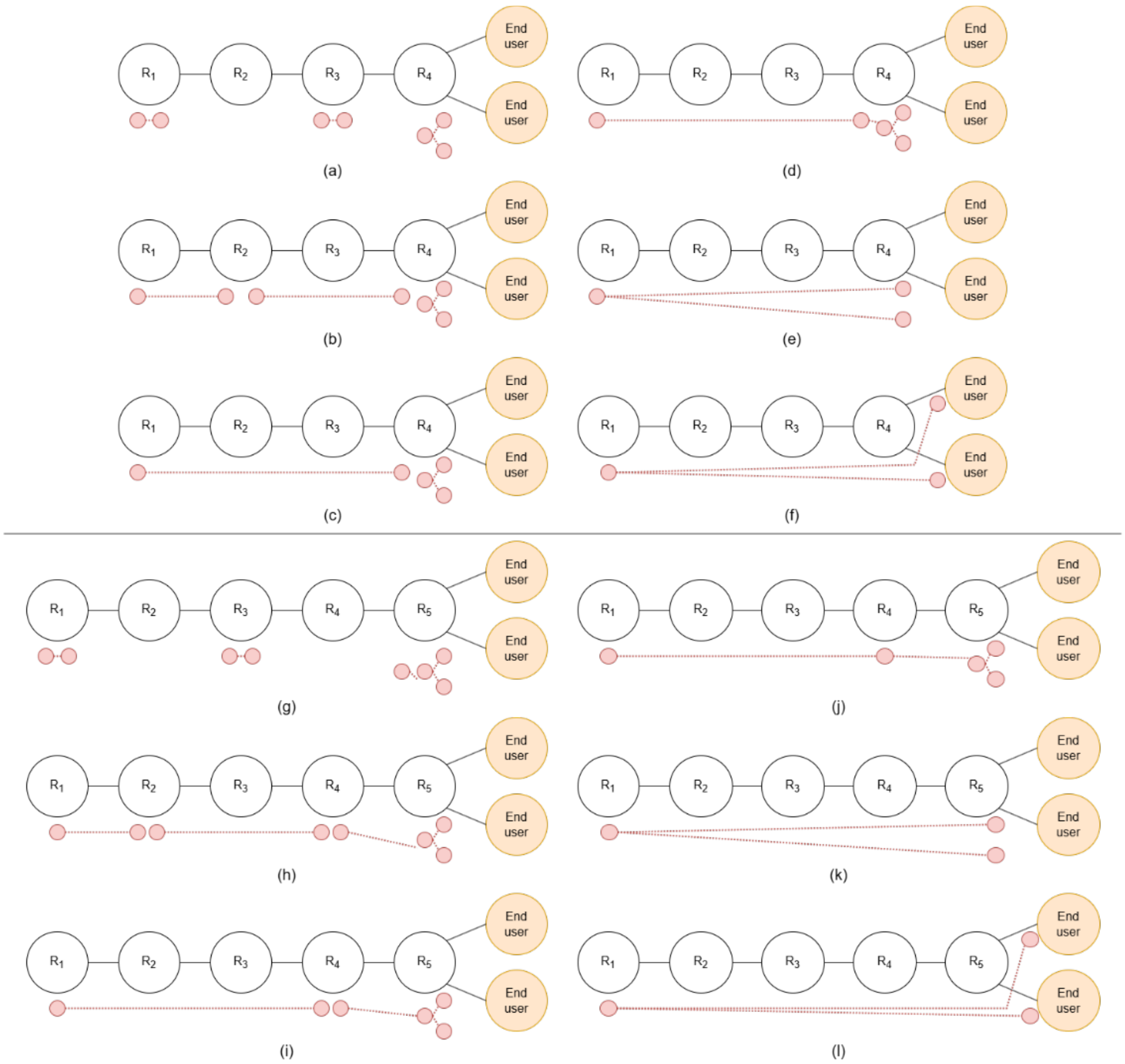

**Fig. 4**. Qubit distribution method, the red circles represent qubits, (a)-(f) and (g)-(l) illustrate the examples with four and five nodes in the repeater chain, respectively. (a) Initial states generated by repeaters. (b) States after qubits are transmitted to neighboring repeaters. (c) Y-basis measurements on qubits held by $r_2$ to remove unused vertices and edges. (d) Establishment of states with intermediate qubits. (e) Resulting state after Y-basis measurements. (f) final state after performing corrections and sending qubits to the corresponding end users. (g)-(i) Same as (a)-(c). (j) CZ operation on both qubits held by $r_4$, then Y-basis measurement on the former qubit. (k) Sequentially measure qubits held by $r_5$ and $r_4$. (l) Same as (f).

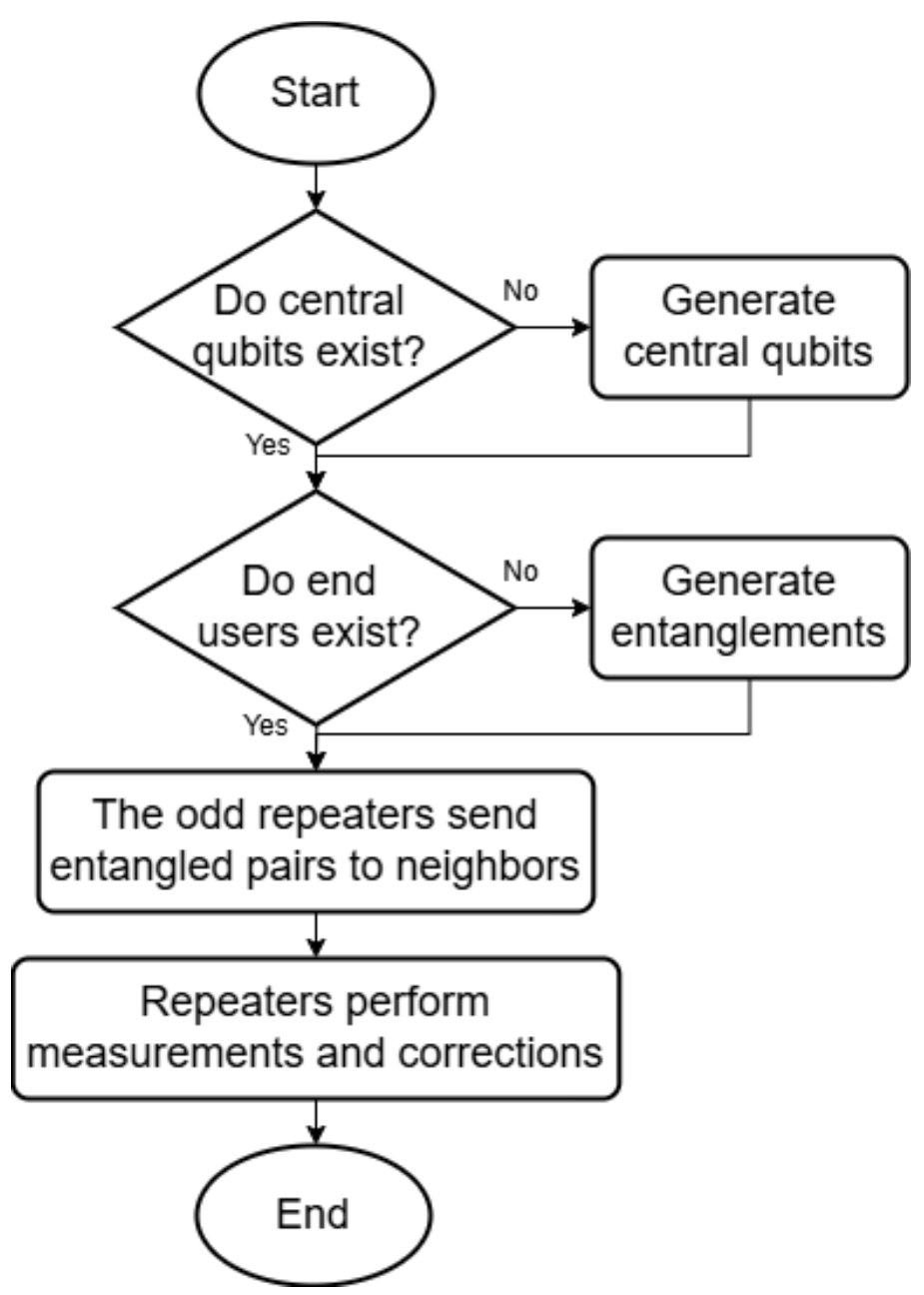

**Fig. 5**. A brief flow chart for the process.

### 3.2 Proposed Distribution Scheme

Given an arbitrary quantum network topology with parameters $n, m$, and $k$, representing the number of nodes (quantum repeaters), edges, and terminal nodes, respectively. In this study, a terminal node is defined as a repeater connected to at least one protocol participant. The objective of the proposed scheme is to efficiently distribute a star graph state among protocol participants with the assistance of quantum repeaters, while minimizing decoherence effects during transmission. As demonstrated in **Section 3.1**, any two star graph states can be merged into a larger star graph state through a sequence of Y-basis measurements and corresponding local correction operations. By incorporating this merging method into the distribution scheme, entanglement can be established among multiple distant participants while ensuring that each intermediate edge is traversed only once, thereby reducing the total number of qubits consumed and minimizing decoherence.

#### 3.2.1 Routing Method

To formalize this optimization, we model the quantum network as a weighted graph $G = (V, E)$, where nodes correspond to repeaters and edges represent quantum channels. The problem of selecting the optimal set of edges for distributing entanglement among the terminal nodes can be naturally formulated as the minimum Steiner tree (MST) problem [32]. The MST connects all terminal nodes with minimal total cost, translating to reduced transmission time and decoherence, thus improving the fidelity of the distributed target star graph state. Specifically, given edge weights $w: E \rightarrow \mathbb{R}^+$ and a subset of terminal nodes $T \subseteq V$, where $\mathbb{R}^+ = \{x \in \mathbb{R} | x > 0\}$, we can leverage state-compressed dynamic programming (DP) to solve the problem. Let $f(i, S)$ denote the minimum edge weight of a tree rooted at node $i$ that spans all nodes in the set $S$. The DP recursion proceeds with two transitions shown in **Eqs. (7)** and **(8)**, respectively. **Eq. (7)** represents the transition between connected subsets, whereas **Eq. (8)** allows for edge relaxation, updating the connection state within the current subset.

$$f(i,S) \leftarrow \min_{T \subseteq S}\big(f(i,S), f(i,T) + f(i,S-T)\big) \tag{7}$$

$$f(i,S) \leftarrow \min\big((f(i,S), f(i,T) + w(j,i)\big) \tag{8}$$

Using this DP approach, an optimized MST can be computed for the quantum network. However, because finding an MST is an NP-hard problem, the above method becomes computationally infeasible for large-scale quantum networks. Therefore, we can consider approximate solutions such as the LP-based approximation approach [33] or the methods presented in [34,355]. In this work, we adopt the approach introduced in [35].

In the experiments conducted in this study, for simplicity, the weight of an edge $e_{ij} \in E$ is defined as the physical length of the quantum channel connecting repeaters $v_i$ and $v_j$:

$$w_{ij} = d_{ij} = \sqrt{\left(x_i - x_j\right)^2 + \left(y_i - y_j\right)^2}, \tag{9}$$

where $(x_i, y_i)$ and $(x_j, y_j)$ are the coordinates of the corresponding repeaters.

Accordingly, the objective of Steiner tree here is to minimize the total lengths of the selected channels, therefore reducing the total physical transmission distance. Assuming a fixed propagation speed, it also reduces the aggregate transmission time and the duration for which transmitted qubits are exposed to channel noise. In the present simulations, dephasing and depolarization parameters are uniform across all edges within each experimental configuration and are not included directly in the edge weights.

It should be noted that in a practical quantum network, however, the edge weight should also account for heterogeneous channel noise, entanglement-generation success probability, transmission latency, quantum-memory availability, link workload, and local resource reliability. These factors are not included in the current experiments and are left for future work.

### 3.2.2 Selection on Root in MST

After an MST is established, the total edge weight is minimized. However, the selection on root in MST is still a dominant factor in determining overall resources consumption and performance. Therefore, here we provide two candidates to be the root in MST:

(1) TP in the QCP: If the purpose of the distribution of target states is for an execution for QCP, and that QCP has a TP to help the execution, then we can select that TP as the root. Therefore, we don't have to consume more resources after the distribution to teleport central qubit to the TP.

(2) Central nodes of star graph states: A central node of a star graph state must be the beginning or ending of an intermediate qubit chain connecting two star graph states. If we select qubit in the middle of a qubit chain, the qubit chain will be broken into two chains and thus consume more qubits and require more operations. Therefore, the selection on central nodes will be a better choice than other qubits.

If multiple candidates satisfy the requirements, we can simply select the one that can minimize the maximum distance to other network nodes, thus minimize the communication latency to speed up the execution of the protocol, thereby maintaining higher fidelity of final states.

### 3.2.3 Prototype of Scheme

The merging of star graph states can be performed in a hierarchical manner. Specifically, if a node has an out-degree greater than one, where out-degree refers to the number of edges leaving a vertex $v$, it temporarily serves as the central node of a star graph state. The construction proceeds from the leaf nodes upward: star graph states are first established at the terminal nodes, and once a star graph is completed at a given node, that node is treated as a single "virtual node" at the next hierarchical level. This process is repeated iteratively until only a single, global star graph state remains.

At each hierarchical level, the construction of a star graph state is accomplished by merging two star graph states connected via a path state, as described in **Section 3.1**. The execution order of these path-based merging operations can be determined using a depth-first search (DFS) [36] traversal of the network graph. Specifically, the order of paths is first established, and for each path, the method introduced in **Section 3.1** is applied. This process is iteratively performed for each path until only a single star graph state remains, after which the leaf qubits are transmitted to the corresponding protocol participants. This hierarchical approach offers an organized and scalable method for constructing large star graph states across arbitrary network topologies.

As illustrated in **Fig. 6**, where the nodes colored orange are the participants, suppose we have built the MST rooted at $R_3$ (**Fig. 6a**), and determined the execution order of paths. We first deal with the paths $\{R_7, P_3\}$ and $\{R_7, P_4\}$, resulting in a star graph state of size 3 and stored at $R_7$ (**Fig. 6b**). Afterward, we process the path $\{R_3, R_4, R_5, R_6, R_7\}$, transferring the entanglements to $R_3$ (**Fig. 6c**). We then iteratively apply the same method to paths $\{R_{10}, P_5\}$, $\{R_3, R_8, R_9, R_{10}\}$, $\{R_1, P_1\}$, $\{R_3, R_2, R_1\}$, and $\{R_3, P_2\}$, as shown in **Figs. 6d, 6e, 6f**. After applying the corresponding correction to the remaining qubits, the leaf qubits are then sent to the participants, finishing the distribution of the star graph state.

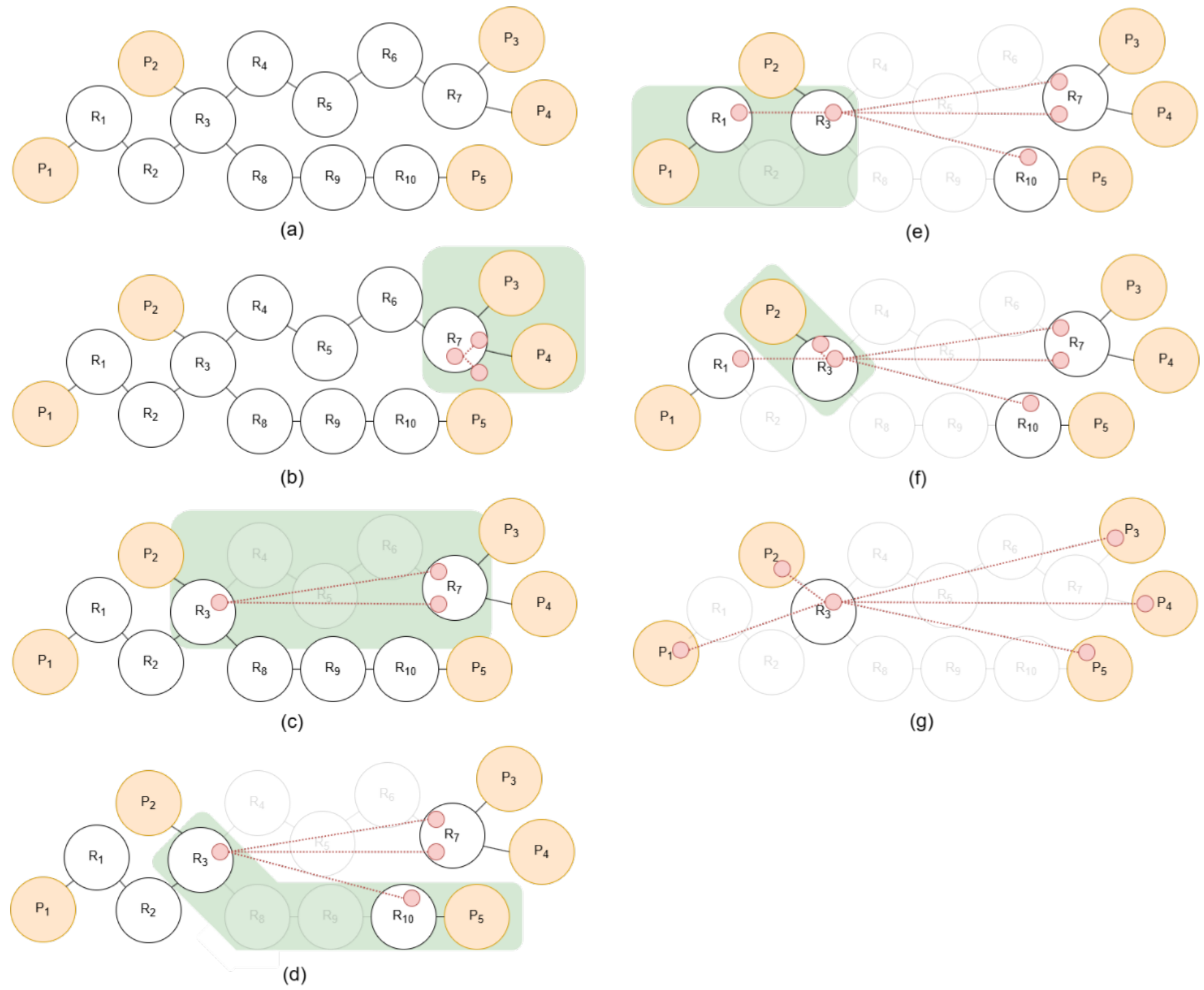

**Fig. 6**. Example of star graph distribution.

### 3.2.4 Proposed Distribution Scheme

We have demonstrated the transformation using the prototype scheme. However, this prototype requires qubits to be measured sequentially, resulting in complex message exchanges and reduced efficiency. Moreover, it necessitates storing qubits for extended periods, which is infeasible with current technology and severely degrades fidelity. To address these challenges, we leverage the operation reductions introduced in **Section 2.3**.

In the prototype, the operations required for entanglement transfer include Y-basis measurements and rotations around the Z-axis. By examining the commutation relations summarized in **Table 1**, we can conclude two key properties: 1). A Y-basis measurement can be replaced with a X-basis measurement combined with classical post-processing if $R_z\left(-\frac{\pi}{2}\right)$ or $R_z\left(\frac{\pi}{2}\right)$ is required to be applied beforehand. 2). The measurement result should be inverted if $R_z\left(-\frac{\pi}{2}\right)$ or $R_z(\pi)$ is required to be applied beforehand. Based on these two properties, we can improve our distribution scheme by replacing sequential measurements with parallel measurements.

Specifically, instead of iteratively performing measurements on qubits during execution, we can perform a "virtual" simulation of the process to determine the dependencies of qubits. This simulation allows us to precompute the appropriate measurement basis of each intermediate qubit.

Once these measurement bases are determined, all intermediate qubits can be measured in parallel after the entanglement distribution process, and the measurement results can then be corrected through classical post-processing according to the computed dependencies. In the original prototype, sequential processing was required because corrections had to be applied to neighboring qubits after each measurement. However, the two properties outlined above allow us to shift the correction step to classical post-processing. This works because the first property ensures that the measurement basis of each qubit depends solely on the number of neighboring qubits measured before it in the virtual process. Once the corrected measurement results of its dependent qubits are known, we can determine the corresponding possible rotation angle to be applied to the qubit. With this information, we can subsequently decide whether the measurement result should be inverted or not. The whole process of graph state distribution is summarized in **Algorithm 1**. The overall time complexity of **Algorithm 1** is dominated by the construction of the (approximate) MST, since all subsequent steps require only $O(N)$ time w.r.t the number of repeaters. The time complexity approximating an MST depends on the chosen algorithm and desired approximation ratio. A commonly used 2-approximation algorithm runs in $O(|V||T|^2)$ [35], where $V$ and $T$ are the vertex set and terminal set, respectively. More sophisticated LP-based approximation schemes can achieve better ratio but at substantially higher computation cost.

---

**Algorithm 1:** Distribution Scheme for Star Graph States

---

1: **Input:** A quantum network graph $G = (V, E)$ with edge weights $w: E \rightarrow \mathbb{R}^+$, a set of protocol participants.
2: **Output:** A distributed star graph state among all protocol participants.
3: Compute an approximate MST $T^* \subseteq E$ that spans all terminal nodes.
4: Select a node in $T^*$ as the central node of the final star graph.
5: Perform DFS to build paths.
6: Virtually perform path reductions to get the dependency and measurement basis for each qubit, as described in **Section. 3.2.4**.
7: Build entangled state among nodes in $T^*$, following the method proposed in **Section. 3.2** but without measurements and corrections.
8: Measure intermediate qubits and announce the measurement results.
9: Perform corrections on the remaining qubits according to the dependencies and measurement results.
10: Sends to qubits from repeaters to protocol participants.
11: **Return:** A distributed star graph shared between protocol participants.

---

We take the same topology illustrated in **Fig. 6** as an example. The entangled state is first established across the network, as shown in **Fig. 7**, where each circle represents a qubit. Circles colored red denote intermediate qubits that will be measured, while circles colored orange indicate qubits that belong to the target star graph state. In the virtual simulation, after each terminal node locally establishes its star graph state, we begin with the path $\{R_3, R_4, R_6, R_7\}$, which corresponds to qubits $\{q_4, q_5, q_6, q_7, q_8, q_9\}$. The qubit $q_5$ is processed. Once it is measured, corrections must be made on $q_4$ and $q_6$, meaning that $q_4$ and $q_6$ are dependent on $q_5$. Next, $q_6$ is processed, adding dependencies to $q_4$ and $q_7$. The process continues iteratively with $q_7$, $q_9$ and $q_8$, with dependencies recorded accordingly. This process is applied to each path in the same order as described in **Section 3.2**, and the resulting dependencies are summarized in **Table 2**. For illustration purposes, the measurement results used in this example are also presented in **Table 2**. After determining the measurement bases, all corresponding measurements are performed on qubits in parallel, obtaining the measurement results, and corrections can then be applied based on the dependency.

For example, consider $q_2$: The actual measurement results of its dependent qubits are both $|-i\rangle$, resulting in a total rotation angle of $\pi$. According to the second property discussed in this section, the measurement result must be inverted, yielding a final result of $|-\rangle$. The corresponding corrections for the remaining qubits can be inferred in the same way, and the order of inferring can be determined by using algorithms like topology sort. After all corrections are applied, leaf qubits are sent to the respective protocol participants.

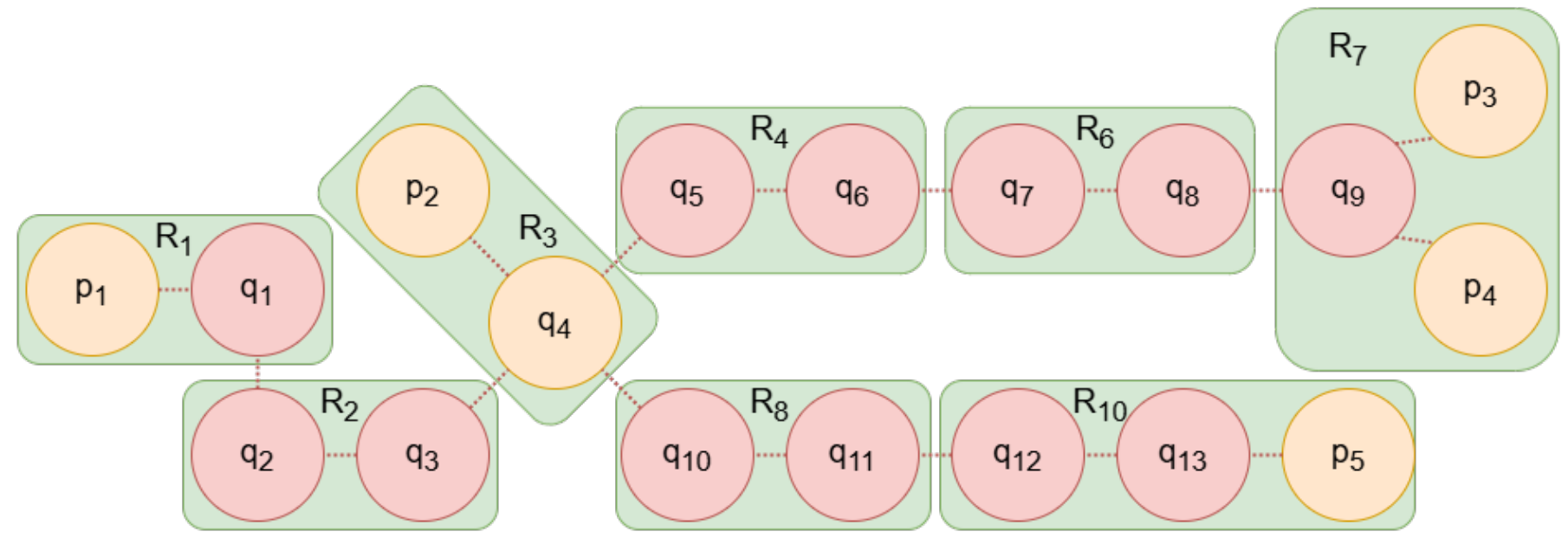

**Fig. 7**. Example entangled state established across quantum networks.

**Table 2**. Dependence summary

| Qubit | Dependent qubits | Possible rotation angle | Quantum Operation | Measurement results | Actual results / operation |
|---|---|---|---|---|---|
| $q_1$ | None | 0 | Y-basis measurement | $\|-i\rangle$ | $\|-i\rangle$ |
| $q_2$ | $q_1, q_3$ | 0 or $\pi$ | X-basis measurement | $\|+\rangle$ | $\|-\rangle$ |
| $q_3$ | None | 0 | Y-basis measurement | $\|-i\rangle$ | $\|-i\rangle$ |
| $q_4$ | $q_2, q_3, q_5, q_6, q_7, q_8, q_{10}, q_{11}, q_{12}$ | $\pi/2$ or $-\pi/2$ | $R_z(\pi/2)$ or $R_z(-\pi/2)$ | None | $R_z(-\pi/2)$ |
| $q_5$ | None | 0 | Y-basis measurement | $\|i\rangle$ | $\|i\rangle$ |
| $q_6$ | $q_5$ | $\pi/2$ or $-\pi/2$ | X-basis measurement | $\|+\rangle$ | $\|-\rangle$ |
| $q_7$ | $q_6$ | $\pi/2$ or $-\pi/2$ | X-basis measurement | $\|-\rangle$ | $\|-\rangle$ |
| $q_8$ | $q_7, q_9$ | 0 or $\pi$ | Y-basis measurement | $\|-i\rangle$ | $\|i\rangle$ |

| $q_9$ | None | 0 | Y-basis measurement | $\|i\rangle$ | $\|i\rangle$ |
|---|---|---|---|---|---|
| $q_{10}$ | None | 0 | Y-basis measurement | $\|-i\rangle$ | $\|-i\rangle$ |
| $q_{11}$ | $q_{10}$ | $\pi/2$ or $-\pi/2$ | X-basis measurement | $\|+\rangle$ | $\|+\rangle$ |
| $q_{12}$ | $q_{11}, q_{13}$ | 0 or $\pi$ | Y-basis measurement | $\|i\rangle$ | $\|i\rangle$ |
| $q_{13}$ | None | 0 | Y-basis measurement | $\|i\rangle$ | $\|i\rangle$ |
| $p_1$ | $q_1, q_2$ | 0 or $\pi$ | $I$ or $R_z(\pi)$ | None | $I$ |
| $p_2$ | None | 0 | $I$ | None | $I$ |
| $p_3$ | $q_8, q_9$ | 0 or $\pi$ | $I$ or $R_z(\pi)$ | None | $R_z(\pi)$ |
| $p_4$ | $q_8, q_9$ | 0 or $\pi$ | $I$ or $R_z(\pi)$ | None | $R_z(\pi)$ |
| $p_5$ | $q_{12}, q_{13}$ | 0 or $\pi$ | $I$ or $R_z(\pi)$ | None | $R_z(\pi)$ |

## 4. Simulation, Comparison, and Evaluation

This section first presents the simulation setup and results, followed by a comparison with related work, and evaluates the feasibility for real-world implementation.

### 4.1 Simulation Setup

This study implements the proposed scheme using a quantum network simulation tool, NetSquid [28], to verify its correctness and evaluate the impact of quantum channel noise. The simulation experiments comprise three main parts:

(1) Performance evaluation under varying network conditions: We assess how network parameters, such as the number of quantum repeaters, channel noise, and edge density, affect the performance of the proposed scheme.
(2) Feasibility analysis in application scenarios: We estimate the feasibility of the proposed scheme by implementing and comparing the QCP from [37] under two different settings: (a) protocol participants have direct quantum channels to the third party (TP); and (b) protocol participants rely on quantum networks to distribute the entangled state.
(3) Impact of root selection and MST optimality: We assess the impact of root selection and MST optimality on the final distribution success rate.

In this study, “transmission” refers specifically to the propagation of a qubit through a quantum channel between two neighboring repeaters. Time-dependent dephasing begins when the qubit is sent and ends when it arrives at the receiving repeater; local gates and measurements are assumed to be instantaneous, while quantum-memory waiting time and classical-communication delay are excluded from the current timing model. Additionally, As this study is primarily focused on transmission, we do not consider the memory lifetime and operation error rates to give a concentrate analyze on decoherence without interfering by other factors. Although the memory lifetime and operation error rates are important, they are hardware-specific and outside the scope of this study. Note that the memory lifetime of the proposed method is low because of the parallel processes. For the channel noises, we consider the following two provided by Netsquid:

(1) Dephasing noise: This noise models the loss of quantum phase coherence without affecting populating probabilities. In quantum networks, dephasing arises from environmental fluctuations, optical-path instability, and imperfect synchronization, and it primarily degrades entangled states by reducing off-diagonal terms of the density matrix. In this study, we model dephasing using a time-dependent dephasing rate with a default rate of 500 Hz, varying from 100 to 1500 Hz.

(2) Depolarization noise: Depolarization noise describes a uniform random Pauli error channel in which a qubit's state is replaced by a maximally mixed state with probability $p$. This noise captures imperfections from photon loss, scattering, and faulty gate operations. This mechanism affects both phase and amplitude components, making it a widely used approximation for device-level noise in quantum networks. In this study, we model depolarization using a time-independent depolarization probability with a default value of 0.1, ranging from 0.01 to 0.25.

In the following, we denote channel noise as $(dephasing\ rate,\ depolarization\ rate)$. We set the values of channel noise and other simulation parameters introduced below by following the recent works [16,19,38,39].

- The parameter set of 1st simulation

In the first set of simulations, the quantum networks are generated randomly by using the Waxman model [40]. Nodes are distributed over a $100\ km \times 100\ km$ area and vary the Waxman parameters $\alpha$ and $\beta$ to adjust the edge density according to the network size. The default edge density is 0.1, and we explore the values in $\{0.05, 0.1, 0.15, 0.2\}$. The number of quantum repeaters is 100 by default, and ranges from 10 to 150, while the number of terminals is 9 by default, and varies from 7 to 20. For simplicity, we assume each terminal corresponds to a unique protocol participant. The channel length here is equal to Euclidean distance between two nodes. All MSTs are selected using the method introduced in [35], and roots are randomly chosen.

For each data point, 500 simulation runs are performed, and in each run, a star graph state is distributed across the quantum network. Upon completion, the entangled state is measured to verify successful distribution by checking the properties described in [5]. The success rate is calculated as the ratio of successful distributions to the total number of simulation runs.

- The parameter set of 2nd simulation

In the second set of simulations, we implement the QCP proposed in [37] under two different scenarios to assess the feasibility of the proposed scheme in practical settings. In the first scenario, protocol participants are assumed to be directly connected to TP via individual quantum channels. The lengths of these channels range from 20 km to 40 km, and three different noise configurations, $\{(500,0), (0,0.1), (500,0.1)\}$, are considered to represent pure dephasing, pure depolarizing, and mixed noise effects, respectively.

In the second scenario, participants may not be directly connected to TP and must rely on an intermediate quantum network to distribute the entangled state. The configurations of the quantum networks in this scenario follow the same setup as in the first set of simulations. Additionally, we assume that the root of MST is also the TP of QCP.

The success rate in each simulation run is calculated as in [37], defined as the ratio of correctly shared bits to the total number of secret bits. Each data point is obtained from 10 simulation runs, with the lengths of both secret bits and check bits set to 256 in each run. The final success rate is derived by averaging the results over the 10 simulation runs.

It should be noted that the QCP in [37] adopts a centralized structure in which a TP coordinates the protocol participants. This structure is naturally compatible with the star graph states generated by the proposed scheme. Therefore, the experiment is intended to evaluate the feasibility of the proposed method for its target class of applications, rather than to provide a topology-neutral comparison with general-purpose graph-state distribution schemes.

- The parameter set of 3rd simulation

In the third set of simulations, we conduct an ablation study to evaluate the respective impact of MST optimality and root selection on the performance of the proposed scheme. The configurations of the quantum networks and assessments of success rates follow the same setup as in the first set of simulations, and each simulation is performed under noise setting $(500,0.1)$.

To decouple the effects of tree construction and root placement, four configurations are considered:

(1) Optimal MST with optimal root.
(2) Optimal MST with random root.
(3) Approximate MST with optimal root.
(4) Approximate MST with random root.

The optimal and approximate MSTs are constructed using the method described in **Section. 3.2.1** and [35], respectively, while the optimal root is selected according to the strategy introduced in **Section. 3.2.2**.

### 4.2 Simulation Results

The results for the three sets of simulations are presented in **Figs. 8, 9** and **10**. In **Fig. 8a**, it can be observed that as the number of terminals increases, the success rate steadily decreases under all noise configurations, with combined noise significantly reducing the success rate compared to single-noise settings. In **Fig. 8b**, where the number of terminals is fixed at 9, increasing the number of nodes in the quantum network has a limited effect on the success rate; a larger network does not necessarily improve performance without enhanced connectivity or greater noise resilience. In **Fig. 8c**, the numbers of nodes and terminals are fixed while the noise settings are varied. The results show that both depolarization and dephasing rates significantly degrade the success rate, especially when combined. This is consistent with previous simulations, indicating that the scheme tolerates only one high-noise source at a time. In **Fig. 8d**, the edge density varies from 0.05 to 0.2; the results indicate that higher edge density improves performance, particularly in the presence of noise. With the noise configuration (500, 0.1), the success rate doubles as the edge density increases from 0.05 to 0.2. This demonstrates the advantage of redundant paths and improved connectivity in mitigating quantum channel noise.

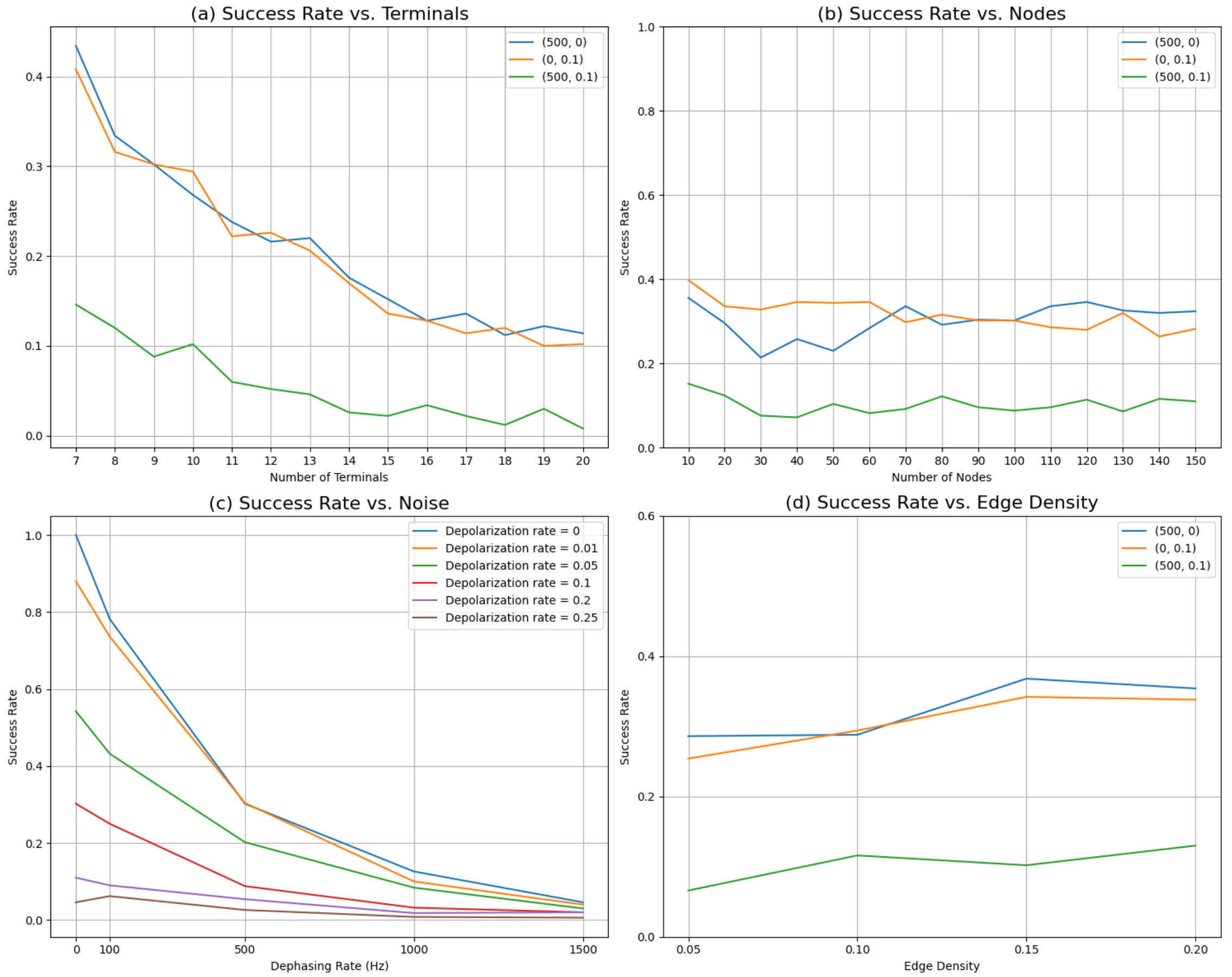

**Fig. 8**. Simulation results for the first set of simulations.

**Fig. 9** illustrates the performance of the QCP proposed in [377] under different application scenarios. Lines 1 to 5 represent the case in which each protocol participant has a direct quantum channel to TP, while lines 6 to 8 correspond to scenarios where participants are randomly selected from a quantum network and the entangled states are distributed with the assistance of intermediate quantum repeaters. The overall lower success rates of lines 6 to 8, compared to lines 1 to 5, reflect the added complexity and increased vulnerability to noise resulting from entanglement distribution across quantum networks.

Furthermore, the performance trend of our scheme closely aligns with the assumptions in [37], indicating its feasibility and correctness. Notably, the results also demonstrate the sensitivity of entanglement distribution and QCPs to both channel noise and transmission distance, highlighting the importance of network optimization and the development of noise-resilient protocols.

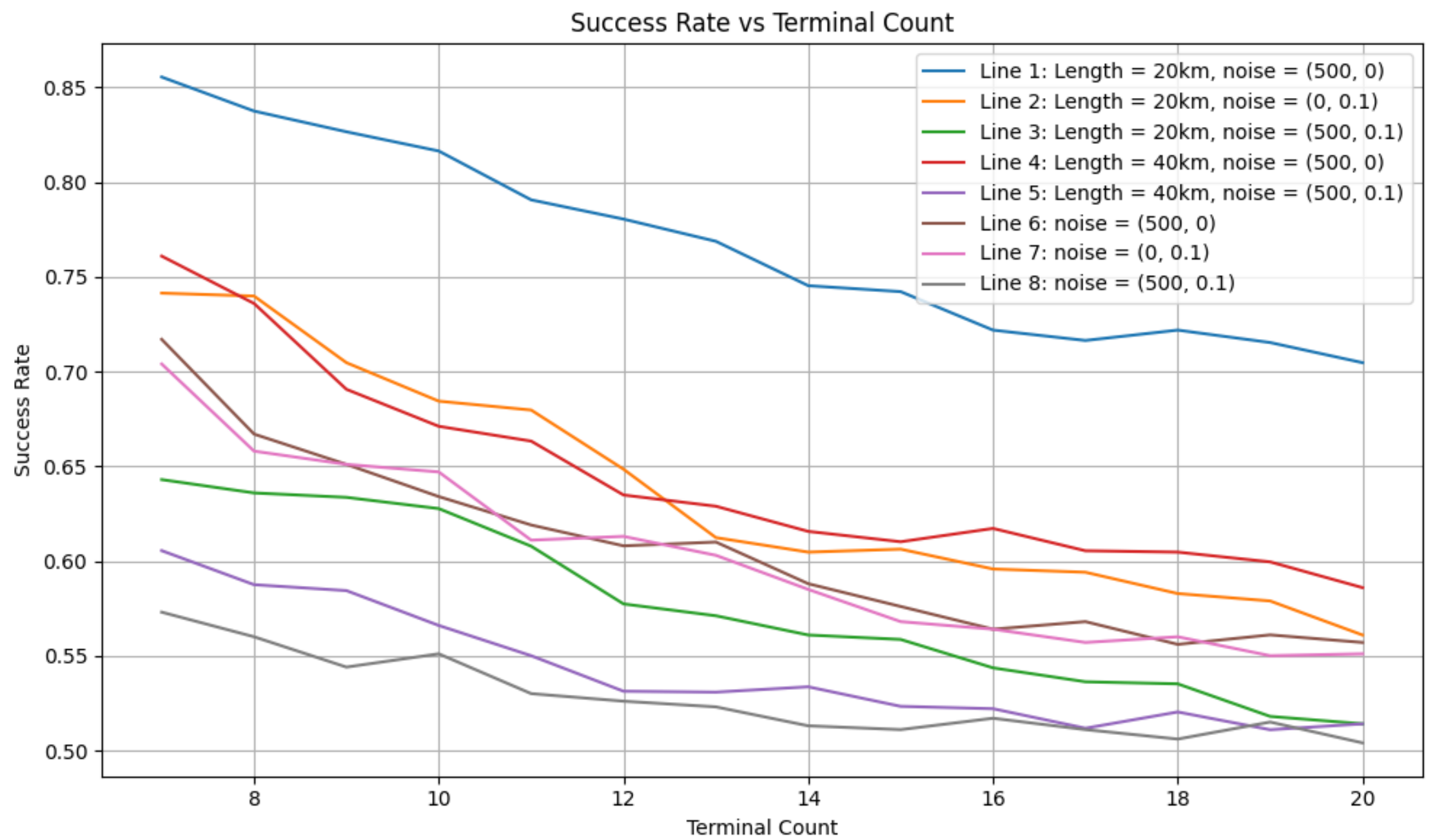

**Fig. 9**. Simulation results for the second set of simulations.

**Fig. 10** presents an ablation study to evaluate the respective impacts of MST optimality and root selection. **In Fig. 10a**, the success rate is shown as the number of terminals increases. It can be observed that the configuration using optimal MST with optimal root selection consistently achieves the highest across all terminal settings. When the root is randomly selected, the performance degrades noticeably even with the optimal MST, indicating that root placement plays a significant role in the entanglement distribution process. Moreover, employing and approximate MST further reduces the success of rate, and the degradation become more pronounced when combined with random root selection. **Fig. 10b** shows the success rate under varying numbers of network nodes. Similar trends are observed. The optimal MST with optimal root again provides the best performance, while both random root selection and approximate MST construction lead to measurable performance loss. Notably, for a fixed MST type, selecting an appropriate root consistently improves the success rate, demonstrating that root selection affects the effective transmission depth and decoherence exposure beyond the tree structure alone.

Overall, the results in **Fig. 10** confirm that both the quality of the MST construction and the root selection contribute to the performance of the proposed distribution scheme. While the MST determines the global communication cost, proper root selection further improves the success probability by reducing the effective propagation depth of entanglement.

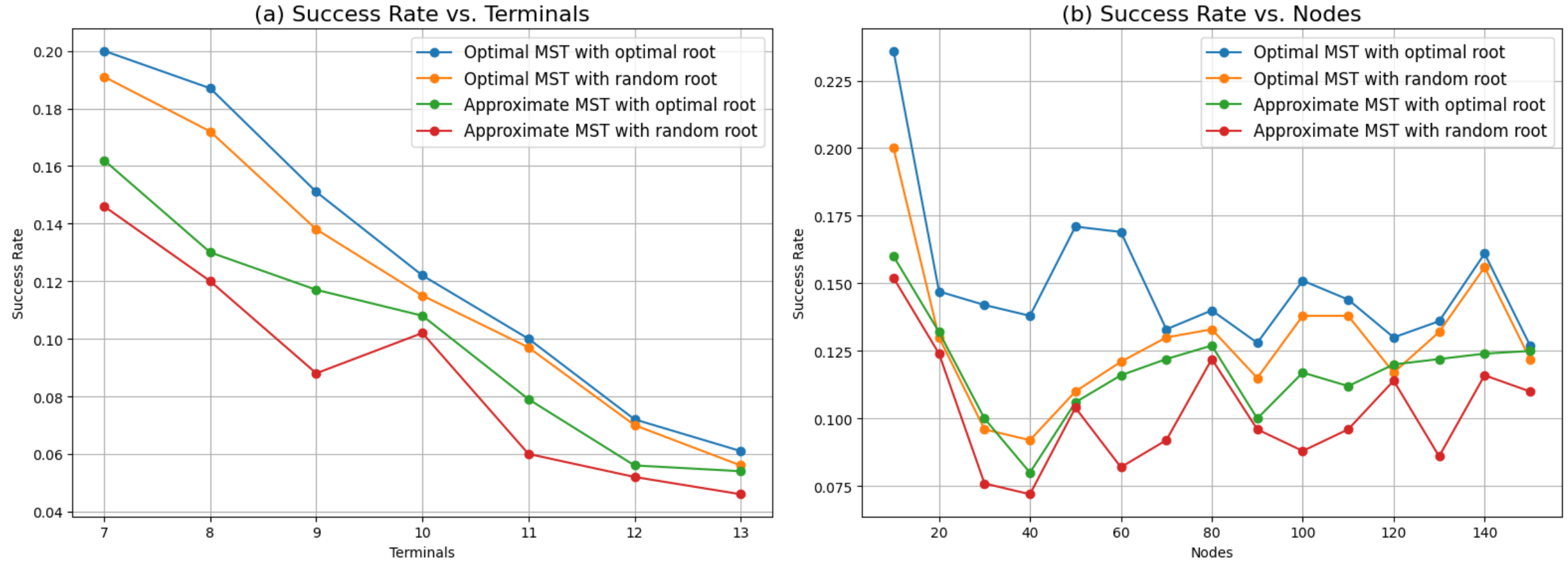

**Fig. 10**. Simulation results for the third set of simulations.

### 4.3 Comparisons and Evaluations

In this section, we provide a detailed evaluation of the proposed scheme from two complementary perspectives. First, we compare the proposed merging strategy based on the graph state operations with BSM approach in [41] to illustrate its effectiveness in constructing multipartite entangled state. Second, we investigate the performance of the proposed distribution scheme with previous studies. These comparisons aim to clarify the contribution of each component in the proposed scheme and to demonstrate its applicability in realistic quantum network environments.

#### 4.3.1 Resource Analysis of the Proposed Merging Method

To evaluate the structural resource requirements of the proposed merging method introduced in **Section 3.1**, we compare it with the multi-user entanglement swapping-based network fusion approach presented in [41]. In [41], two independent fully connected networks are merged by performing a BSM between designated gateway nodes, thereby establishing inter-network entanglement in the quantum correlation layer. It should be emphasized that this comparison focuses on the graph structures and local merging operations, rather than the complete resource consumption of entanglement distribution over a quantum network.

In particular, the numbers reported for the establishment stage describe the number of edges in the initial entangled graph structures. Under an ideal circuit-level graph-state abstraction, each edge corresponds to one entangling operation, such as a CZ gate. However, this count does not include additional CZ operations performed at intermediate repeaters during network-level distribution. Such operations depend on the selected routing tree, the number of repeaters, and the lengths of the transmission paths, their effects are instead reflected in the network-level distribution procedure and simulations described in **Sections 3.2** and **4.2**.

Suppose that each initial network contains $N$ users and one assistant node. In the proposed method, each assistant node serves as the center of a star graph connected to $N$ user qubits. Thus, each star graph contains $N$ edges, and the two initial star graph states contain a total of $2N$ edges. After the input states have been prepared, merging them requires one auxiliary qubit, two additional entangling operations, two single-qubit measurements, and the corresponding correction operations. The resulting state is a star graph state containing all $2N$ user qubits and one remaining central qubit.

In contrast, the method in [41] first establishes two fully connected entanglement structures. A fully connected entanglement structure containing $N + 1$ qubits has $\binom{N+1}{2}$ edges. Therefore, two such graphs contain $2\binom{N+1}{2} = N(N + 1)$ edges. Their inter-network fusion is then performed through auxiliary gateway qubits and a Bell-state measurement. At the circuit level, the BSM can be decomposed into one controlled gate together with single-qubit measurements and corrections.

The resulting resource requirements are summarized in **Table 3**. When only the merging stage is considered and the input graph states are assumed to have already been prepared, both approaches require only a constant number of additional quantum operations w.r.t $N$. The principal difference in their total resource consumption arises during the establishment stage, because a star graph contains only $N$ edges, whereas a fully connected graph contains $N(N + 1)/2$ edges. Thus, the linear scaling of the proposed method is primarily associated with its use of a sparse star graph representation. Since star graph states are locally equivalent to GHZ states and can be transformed into complete graph states through local operations, they provide a resource-efficient representation for the class of QCPs considered in this study. Nevertheless, the two approaches target different entanglement structures and network functions; therefore, **Table 3** should be interpreted as a structural resource comparison rather than a complete network-level performance comparison.

**Table 3**. Comparisons between the proposed merging method and BSM approach.

| | **Proposed method** | **Method in [41]** |
|---|---|---|
| **Pre-shared state** | Two star graph states | Two fully connected entangled networks |
| **Edges in the initial graph structures** | $2N$ | $N(N + 1)$ |
| **Additional qubits in merging state** | 1 auxiliary qubit | Auxiliary gateway qubits required for the BSM |
| **Controlled operations in merging stage** | 2 CZ operations | 1 as part of the BSM decomposition |
| **Measurements in merging stage** | 2 single-qubit measurements | 2 single-qubit measurements |
| **Resulting state** | A star graph state of size $2N + 1$ | A Fully connected graph of size $2N$ |

### 4.3.2 Comparisons and Evaluations on Proposed Distribution Method

Due to the limitations of the quantum network simulator NetSquid, the size of the entangled state that can be simulated is restricted. Consequently, we cannot directly derive the average generation rate and fidelity from the simulation results. Although our prototype is sufficient for evaluating the impact of quantum channel noise during transmission, metrics such as generation rate and fidelity also depend on additional factors, such as quantum memory coherence times and gate errors. Therefore, it would be unfair to compare the performance of our prototype, rather than the complete scheme, directly with other works.

In addition, the performance of entanglement distribution is highly sensitive to the quantum network topology. Therefore, rather than directly comparing numerical simulation results, we conduct a qualitative comparison with previous works [19,22,23] based on the following metrics: the types of target states, strategy, the number of qubits consumed per shot, the number of operations required per shot, and memory management. The results are summarized in **Table 4**.

Our proposed scheme is specifically designed to distribute star and complete graph states, which are highly leveraged by many QCPs. In contrast, [19] and [22] target general graph states, offering more flexibility and versatility but at the cost of higher resource overhead and implementation complexity. On the other hand, [22] focuses on only GHZ states, which are a subset of star graph states, making it less versatile compared to our approach. In terms of operation strategy, our scheme consists of three stages, which is intuitive and simple. [19] requires solving a linear programming (LP) problem, which results in high computation complexity. Both [22] and [23] adopt centralized approaches to distribute target states, consuming more qubits than our scheme. Regarding the qubit consumption per shot, [19] may result in unpredictable qubit usage depending on the LP result and fusion structure. As for our scheme, as well as [22] and [23], all achieve $O(N)$ qubit space complexity w.r.t the number of nodes $n$. However, [22] requires redundant graph structures, and [23] uses even more qubits compared to [22] because of the centralized approach. Therefore, our scheme is more resource-efficient in this metric. As for the required operations, all the methods show high similarity since only controlled-Z (CZ) gates and local unitary operations are required. In the case of [19], the fusion operations can be decomposed by the properties introduced in **Section 2.1**, making them achievable in practical implementation. Finally, w.r.t quantum memory management, our approach avoids using long-term quantum storage by enabling parallel operations, making it suitable for lightweight application environments. In contrast, [19] takes advantage of long-term quantum memories to achieve higher generation rate, but this could lead to high cost to establish such hardware. Neither [22] nor [23] explicitly provides a strategy to optimize memory usage, making them more vulnerable to decoherence over time.

**Table 4**. Comparisons between the proposed scheme and previous works.

| | **Ours** | **[19]** | **[22]** | **[23]** |
|---|---|---|---|---|
| **Target state** | Star and complete graph states | General graph states | General graph states | GHZ states |
| **Strategy** | 1. Find MST.<br>2. Build the entangled state.<br>3. Parallel measurements and corrections. | 1. Compute fusion structure via LP-based approach.<br>2. Use fusion operations to build states. | 1. Construct local entangled states.<br>2. Transfer entanglements. | 1. Build links between the central node and clients.<br>2. Transfer entanglements. |
| **Qubits consumed** | $O(N)$ with selected paths. | Depends on LP and fusion structure. | $O(N)$ with selected paths, needs redundant graphs. | $O(N)$ with selected paths, greater than ours and [20]. |
| **Operations required** | CZ gates,<br>X-/Y-basis measurements. | CZ gates,<br>Pauli-Y/Z gates,<br>Fusion operations. | CZ gates,<br>BSMs,<br>Local Clifford operations. | CZ gates,<br>BSMs,<br>Pauli-Y/Z gates. |
| **Memory management** | Minimized via parallel process, no need for long-term storage. | Modeled in LP via constraints, needs long-term storage. | No explicit optimization. | No explicit optimization. |

Although the generation rate and fidelity cannot be directly evaluated through simulation, it can be inferred that methods which retain entangled states, such as those in [16,18,19], may achieve higher generation rates and reduced total resource consumption by treating these entanglements as reusable intermediate edges. However, this approach results in lower fidelity, as the increased coherence time required for quantum memory raises the risk of decoherence and noise accumulation.

Although this study evaluates the proposed method primarily through simulation, the required quantum operations are achievable in near-term quantum devices. In the proposed scheme, the only multi-qubit operation is the CZ operation, while the remaining operations consist of single-qubit Pauli-basis measurements and Z-axis rotations. These operations belong to the Clifford gate set and have been widely demonstrated with high fidelity. Moreover, the merging method described in **Section 3.1** involves a limited number of controlled operations. In current photonic and superconducting implementations, CZ gates and Pauli measurements are among the most mature and high-fidelity operations [42], supporting the practical feasibility of the proposed scheme.

Regarding fidelity requirements, the dominant factor is the coherence time of qubits during one distribution round. Because the proposed scheme minimizes the duration for which qubits must remain idle, the accumulated noise is thus reduced relative to methods that require long-term qubit storage or sequential multi-step fusion procedures. Therefore, although the absolute fidelity requirement depends on the target application (e.g., QKD, QSS, etc.), the proposed method imposes no stricter constraints than existing approaches and can maintain higher end-to-end fidelity under realistic decoherence rates.

In the real quantum network, a fundamental requirement is the support of multiple concurrent end-to-end communication requests. In the proposed scheme, each distributed GHZ state is dedicated to a single communication request. Therefore, when multiple concurrent requests arise, the proposed distribution procedure must be executed multiple times. In such cases, certain quantum channels may be repeatedly utilized, potentially leading to resource overloading. To address this issue, existing routing and resource allocation strategies, such as those proposed in [16] and [34], can be incorporated. In particular, [16] formulates entanglement constraints as subjections to linear programming (LP) models. Following this approach, the proposed method can accommodate concurrent requests by appropriately adjusting the edge weights in the network graph, where the weights reflect channel conditions, such as workload and noise level. The specific formulation of edge weights can be flexibly designed according to system requirements and application scenarios. In addition, [34] assigns each edge to a dedicated Steiner tree. Since the proposed method is based on MST construction, this strategy can be naturally integrated into our framework to further support multiple concurrent end-to-end communication requests.

## 5. Conclusion

Existing studies on the distribution of graph states pursue various objectives, yet few explicitly address the challenge of decoherence during transmission. In this work, we leverage the properties of graph states to propose a novel distribution scheme for GHZ-equivalent graph states across quantum networks, with a particular focus on analyzing and mitigating the effects of transmission-induced decoherence. Compared to previous approaches, the proposed scheme achieves lower qubit consumption and requires fewer quantum operations. Furthermore, the scheme is implemented and evaluated using the NetSquid simulator, and the simulation results verify its correctness and feasibility for enabling quantum communication protocols over quantum networks.

Future research could explore generalizing the proposed scheme to support other classes of graph states, such as grid and bipartite graph states, which are relevant in measurement-based quantum computation and distributed quantum sensing. In addition, since the current scheme does not utilize long-term quantum memories, investigating the integration of memory-based optimizations, such as those proposed in [17–19], to enhance generation rate and robustness under practical noise constraints represents a promising direction. Furthermore, as this work primarily focuses on resource allocation rather than the full network service stack, mechanisms for supporting multiple concurrent end-to-end requests using distributed entangled states warrant further investigation to improve performance in realistic quantum network scenarios. Finally, developing analytical models to estimate the fidelity and generation rate of the proposed scheme, beyond the limitations of simulation tools like NetSquid, could provide deeper theoretical insights.

## Acknowledgment

This research was partially supported by the National Science and Technology Council, Taiwan, R.O.C. (Grant Nos. 113-2221-E-025 -014 -, 114-2221-E-025 -006 -MY2, and 113-2634-F-005 -001 -MBK).

## Declarations

### Ethical Approval and Consent to participate

Not applicable.

### Consent for publication

Not applicable.

### Availability of supporting data

Data sharing not applicable to this article as no datasets were generated or analyzed during the current study.

### Competing interests

The authors certify that there is no actual or potential conflict of interest in relation to this article.

### Authors' contributions

Conceptualization, C.-W. Tasi and C.-H. Wang; methodology, C.-W. Tasi and C.-H. Wang; investigation, C.-W. Tasi and C.-H. Wang; formal analysis, C.-W. Tasi; writing—original draft, C.-H. Wang; writing—review & editing, C.-W. Tsai; project Administration, C.-W. Tasi. All authors have read and agreed to the published version of the manuscript.

**Appendix A**. Equivalence of $\sqrt{Z}$ and $\sqrt{Z}^{\dagger}$ to rotation operations

In quantum computing, gates such as $\sqrt{Z}$ and $\sqrt{Z}^{\dagger}$ (as mentioned in **Section. 2.3**) can be expressed as rotation operations on the Bloch sphere. Here, we give the proof of equivalence of $\sqrt{Z}$ and $\sqrt{Z}^{\dagger}$ to $R_z\left(\frac{\pi}{2}\right)$ and $R_z\left(\frac{-\pi}{2}\right)$, respectively.

The rotation around Z-axis by $\theta$ can be expressed as:

$$R_z(\theta) = e^{-i\frac{\theta}{2}Z} = \cos\frac{\theta}{2}I - i\sin\frac{\theta}{2}Z. \tag{9}$$

Following this, we can derive the matrix forms of $R_z\left(\frac{\pi}{2}\right)$ and $R_z\left(\frac{-\pi}{2}\right)$:

$$R_z\left(\frac{\pi}{2}\right) = \cos\frac{\pi}{4}I - i\sin\frac{\pi}{4}Z = \frac{1}{\sqrt{2}}\begin{bmatrix}1-i & 0\\ 0 & 1+i\end{bmatrix} \tag{10}$$

$$R_z\left(\frac{-\pi}{2}\right) = \cos\frac{-\pi}{4}I - i\sin\frac{-\pi}{4}Z = -\frac{1}{\sqrt{2}}\begin{bmatrix}1+i & 0\\ 0 & 1-i\end{bmatrix} \tag{11}$$

By adding a global phase $\frac{1-i}{\sqrt{2}}$, we can transform $\sqrt{Z} = \begin{bmatrix}1 & 0\\ 0 & i\end{bmatrix}$ to $R_z\left(\frac{\pi}{2}\right)$:

$$\frac{1-i}{\sqrt{2}}\begin{bmatrix}1 & 0\\ 0 & i\end{bmatrix} = \frac{1}{\sqrt{2}}\begin{bmatrix}1-i & 0\\ 0 & i-i^2\end{bmatrix} = \frac{1}{\sqrt{2}}\begin{bmatrix}1-i & 0\\ 0 & 1+i\end{bmatrix} = R_z\left(\frac{\pi}{2}\right) \tag{12}$$

In the same manner, we can transform $\sqrt{Z}^{\dagger} = \begin{bmatrix}1 & 0\\ 0 & -i\end{bmatrix}$ to $R_z\left(\frac{-\pi}{2}\right)$ by adding a global phase $-\frac{1+i}{\sqrt{2}}$:

$$-\frac{1+i}{\sqrt{2}}\begin{bmatrix}1 & 0\\ 0 & -i\end{bmatrix} = -\frac{1}{\sqrt{2}}\begin{bmatrix}1+i & 0\\ 0 & -i-i^2\end{bmatrix} = -\frac{1}{\sqrt{2}}\begin{bmatrix}1+i & 0\\ 0 & 1-i\end{bmatrix} = R_z\left(\frac{-\pi}{2}\right) \tag{13}$$

As long as the global phases do not affect the measurement results and operations, we can use rotations $R_z\left(\frac{\pi}{2}\right)$ and $R_z\left(\frac{-\pi}{2}\right)$ to replace $\sqrt{Z}$ and $\sqrt{Z}^{\dagger}$ operations, and the equivalences are shown above.

**Appendix B**. Detailed calculation procedures of **Eq. (6)**

According to [13], some measurements that required specific operations to be applied first can be reduced to other measurements. Here, we take the example in **Section. 2.3**, $P_{Y,\pm}\sqrt{Z}^{\dagger} = \sqrt{Z}^{\dagger}P_{X,\mp}$, to show the relationship.

According to **Eq. (5)**, $P$ can be expressed as $\sqrt{Z}P_{Y,\pm}\sqrt{Z}^{\dagger}$, as shown in **Eq. (6)**. Then, we can calculate $\sqrt{Z}(|i\rangle\langle i|)\sqrt{Z}^{\dagger}$ and $\sqrt{Z}(|-i\rangle\langle -i|)\sqrt{Z}^{\dagger}$, respectively.

$$\begin{aligned}\sqrt{Z}(|i\rangle\langle i|)\sqrt{Z}^{\dagger} &= \begin{bmatrix}1 & 0\\0 & i\end{bmatrix}\cdot\frac{1}{2}\begin{bmatrix}1 & -i\\i & 1\end{bmatrix}\cdot\begin{bmatrix}1 & 0\\0 & -i\end{bmatrix}\\ &= \frac{1}{2}\begin{bmatrix}1 & -i\\-1 & i\end{bmatrix}\cdot\begin{bmatrix}1 & 0\\0 & -i\end{bmatrix}\\ &= \frac{1}{2}\begin{bmatrix}1 & -1\\-1 & 1\end{bmatrix} = |-\rangle\langle -| \end{aligned} \tag{14}$$

In the same way, we can derive that $\sqrt{Z}(|-i\rangle\langle -i|)\sqrt{Z}^{\dagger} = |+\rangle\langle +|$, thus proving that $P = P_{X,\mp}$, and $P_{Y,\pm}\sqrt{Z}^{\dagger} = \sqrt{Z}^{\dagger}P_{X,\mp}$.